# Assessing Brazilian agri-food policies: what impact on family farms?

Valdemar J. Wesz Junior[1], Simone Piras[2], Catia Grisa[3], Stefano Ghinoi[4,5]

[1] = Instituto Latino-Americano de Economia, Sociedade e Política, Universidade Federal da Integracao Latino-Americana, Foz do Iguaçu, PR Brazil. Mail: valdemar.junior@unila.edu.br. ORCID: 0000-0002-8154-7088.

[2] = Social, Economic and Geographical Sciences Department, The James Hutton Institute, Aberdeen, UK. Mail: Simone.Piras@hutton.ac.uk. ORCID: 0000-0003-0334-6800.

[3] = Federal University of Rio Grande do Sul, Porto Alegre, RS, Brazil. Mail: catiagrisaufrgs@gmail.com. ORCID: 0000-0001-6685-4875.

[4] = Department of Economics and International Business, University of Greenwich, London, UK. Mail: S.Ghinoi@greenwich.ac.uk. ORCID: 0000-0002-9857-4736.

[5] = Department of Economic and Management, University of Helsinki, Helsinki, Finland.

# Abstract

Since the beginning of the 1990s, Brazil has introduced different policies for increasing agricultural production among family farms, such as the National Program for Strengthening Family Farming (Pronaf), the technical assistance and rural extension programmes (ATER), and seeds distribution. Despite the importance of these policies for the development of family farming, there is a lack of empirical studies investigating their impact on commercialization of food products. By considering household-level data from the 2014 Brazilian National Household Sample Survey, we use propensity score matching techniques accounting for the interaction effects between policies to compare the commercialisation behaviour of recipients with non-recipients. We find that Pronaf has a significant positive impact on family farmers' propensity to engage in commercialisation, and this effect increases if farmers have also access to ATER. Receiving technical assistance alone has a positive effect, but this is mostly limited to smaller farms. In turn, seed distribution appears not to increase commercialization significantly. A well-balanced policy mix could ensure that, in a country subject to the pressure of international food markets, increased commercialisation does not result in reduced food security for rural dwellers.

1. **Introduction**

Until the beginning of the 21st century, family farming was considered part of the problem of food insecurity (FAO, 2014). This assumption was related to the idea that family farmers mainly need to cope with poverty (Ellis, 1998) and have difficulties increasing efficiency and food productivity. However, in recent years family farming has been recognized as a key category for promoting of rural development and environmental sustainability (FAO, 2016; Suess-Reyes and Fuetsch, 2016), reducing hunger (HLPE, 2013; Graeub et al., 2016) and achieving the Sustainable Development Goals (Ortiz et al., 2018).

In line with this growing interest, developing countries have implemented policies with the aim of supporting family farmers (FAO, 2012; Huang et al., 2013; Chen et al., 2014; Villarreal, 2018), Brazil being one of them (Kakwani et al., 2010). Food security has also become an important socio-political issue in this country, resulting in the implementation of policies for increasing food production for domestic supply and improving access to food (Santarelli et al., 2018). Among other policies, family farmers have mainly benefited from the National Program for Strengthening Family Farming (*Programa Nacional de Fortalecimento da Agricultura Familiar* – Pronaf), created to provide them credit at favourable conditions ; the Federal and State policies for technical assistance and rural extension (*Assistência Técnica e Extensão Rural* – ATER), including the National Policy for Technical Assistance and Rural Extension (*Política Nacional de Assistência Técnica e Extensão Rural* – PNATER); and the Federal and State policies for accessing seeds distribution (Seeds), which distributes low-cost seeds to farmers.

The implementation and effectiveness of these policies have been investigated in several studies (Gazolla, 2004; Miná Dias, 2007; Cunha, 2013; Petersen et al., 2013; Grisa et al., 2014; Diesel et al., 2015). However, results are ambiguous; in particular, the effects on family farmers' decision whether producing for self-consumption or for commercialisation. Gazolla (2004) shows that Pronaf has become a powerful tool for stimulating food production for national and international markets, but has reduced production for self-consumption, worsening family farmers' food security. Grisa et al. (2014) point out that Pronaf has mainly supported the production of commodities, while the promotion of sustainable food systems has been stuck in the background.

Thus, Pronaf has been extensively investigated in the family farming literature, and current studies provide divergent views of its effects on family farmers' strategies. In turn, although the ATER and Seeds policies have been particularly important in Brazilian rural development (Diesel et al., 2015), there is a lack of quantitative studies on their effects on family farmers' food commercialisation. Furthermore, to the best of our knowledge, no empirical work has investigated the combined effect of these policies.

Policy evaluation studies are increasingly focusing on the policy mix theme because of the importance of the synergic effects of multiple policies (Howlett and Rayner, 2013; Mavrot et al., 2019). Policies tackling multidimensional issues related to individuals, families, and communities are often implemented simultaneously. In some cases, the agri-food policies mix is able to support family farmers' autonomy by encouraging diversification strategies and promoting sustainable food systems; in other cases, policy mix fosters specialization and conventional systems of agriculture, undermining the family farm autonomy (Buchenrieder, 2007). Given the multidimensional benefits generated by family farms for rural development, it is important to estimate the effects of agri-food public policies on family farmers' decision-making to prevent negative side effects. Since the decision whether to commercialise part of one's own production is key for family farmers, in this paper we try to answer the following research questions:

*RQ1: Does the participation of Brazilian family farms in the agricultural programs (Pronaf, ATER, and Seeds) result in increased propensity to commercialise?*

*RQ2: Does simultaneous participation in more than one program (policy mix) result in even increased propensity to commercialise?*

To answer these questions, we use data from the 2014 National Household Sample Survey (*Pesquisa Nacional por Amostra de Domicílios* – PNAD) of the Brazilian Institute of Geography and Statistics (*Instituto Brasileiro de Geografia e Estatística* – IBGE). This survey contains information on households' participation in different programs supporting family agriculture, households' characteristics, working activities, and earnings. Moreover, the PNAD identifies those households that allocated part of their agricultural production to commercialization. To assess the impact of Pronaf, ATER, and Seeds on family farmers' propensity to allocate some of their production to commercialization, we use Propensity Score Matching, given that participation in the programs and market orientation could be correlated.

The paper is structured as it follows. Section 2 reviews the literature on family farmers' commercialisation choices. The Brazilian policies on family farming are described in Section 3. Section 4 presents our data and methods, while Section 5 illustrates the results of our model and Section 6 concludes.

2. **Family farming strategies for agricultural products commercialisation**

Agricultural products commercialisation has been extensively discussed in the development literature (e.g. Jaleta et al., 2009; Wiggins et al., 2011). International development institutions, primarily the World Bank, consider this approach an effective strategy to address rural poverty, as it increases productivity, raises rural incomes, and improves food security (World Bank, 2007; Jiren et al., 2020); moreover, it is also considered key to achieve overall economic growth in developing countries (Von Braun and Kennedy, 1994; Von Braun, 1995).

While commercialisation is assumed to be a desirable outcome, this cannot be left to the market alone (Jaleta et al., 2009), and a blend of public and private intervention has been used to foster it (Spielman et al., 2010). Public interventions can be ascribed to two main fields: the promotion of productivity and production for sale, and the linking of farmers to markets (Wiggins et al., 2011). Governments can create an enabling environment by developing rural input and output markets, improving communication and transport infrastructure, providing agricultural extension services, securing property rights, and promoting functioning credit markets (Jaleta et al., 2009). Two key instruments are the provision of credit, and information and training (World Bank, 2007). Loans can be obtained in the credit market, through ad hoc institutions targeting small family farms or be informal (Awunyo-Vitor, 2015), and also it can be provided in kind and be non-repayable (Obayelu, 2016; Sibande et al., 2017). Information is usually provided by extension services, which are also responsible for distributing governmental inputs, like seeds and fertilisers (Kilelu et al., 2014).

Concerning the studies on agricultural commercialisation to date, a first branch of the literature has focused on structural and cultural drivers and barriers. Structural elements include, among others, farm characteristics (Bobojonov et al., 2016), rural infrastructures (Hepp et al., 2019), and capital markets (Vasco et al., 2017). Cultural elements consist in values systems affecting farmer preferences, such as the level of risk aversion (Muriithi and Matz, 2015), market orientation (Abafita et al., 2016), preference for food security over income (Poole et al., 2013), diverse livelihoods objectives (Alexander et al., 2018), or the will to reduce external dependence, which could result in disengagement from formal markets (Varga, 2019). A second, ample branch of the literature has analysed the impact of commercialisation on farmers' income and poverty levels (Schure et al., 2014; Mariyono, 2019). A third one has highlighted the social and ecological benefits generated by semi-subsistence farming, with some authors calling for their recognition by means of policy incentives, for example in the EU CAP (Davidova, 2011). Finally, the critical development literature has studied the impact on rural communities of different agricultural markets (Matenga and Hichaambwa, 2017), underlining themes such as food sovereignty and resilience (Jiren et al., 2020).

Despite the large number of studies focusing on drivers and barriers to commercialisation, research assessing the impact of different policy interventions on commercialisation is limited. This is also due to the challenge of defining "commercialisation" – usually understood as the proportion of agricultural output sold in the market to the total value of agricultural production (Jaleta et al., 2009) – and "commercial farmers" – farmers selling over half of their output (Varga, 2020).

Among the studies linking policy interventions to commercialisation levels, Sinyolo et al. (2019) find that social grants reduce small farmers' propensity to commercialise in South Africa. In turn, Mwangi and Crewett (2019) argue that improved infrastructures (namely irrigation systems) increase the commercialisation of peri-urban agriculture in Kenya. Spielman et al. (2010) review the Ethiopian policies to promote cereal intensification and commercialisation, finding that these have not generated sufficient payoffs. Thus, the outcome is very much context-dependent, and measures that have worked well in a certain region or period are not perfectly reproducible elsewhere or at a different time.

Very few studies have quantitatively investigated the extent to which credit and services provision is used either as an input in market agriculture or to internalise resources through self-provisioning (Hepp et al., 2019). Sibande et al. (2017) find a positive effect of subsidisation of fertiliser prices on commercialisation of maize but no effect of family farmers' self-sufficiency in Malawi. Moving towards transition countries, Kostov et al. (2020) finds that direct payments to farmers in Kosovo have a positive effect on market participation for some products (fruits and vegetables, cereals, and oilseeds), but not in the case of livestock products. These works test the impact a single policy, while interaction effects between policies are not considered.

The Brazilian system of agricultural support is a hybrid one between the promotion of business agriculture and family farm – as also shown by the presence, until 2016, of two ministries (the Ministry of Agrarian Development – *Ministério do Desenvolvimento Agrário* (MDA) and the Ministry of Agriculture, Livestock, and Supply – *Ministério da Agricultura, Pecuária e Abastecimento* (MAPA)). Therefore, this country repre-sents an ideal setting to test if agricultural support policies have promoted a more commercial orientation of family farmers. Schneider and Niederle (2010) argue that the attempts to integrate family farmers in market processes in Brazil generate heterogeneous responses, as the rationale behind family farmers' choices is the search for "autonomy" in the context of increasing vulnerability. Therefore, an alternative outcome to increased commercialisation could be an internalisation of subsidies, with family farmers using them to consolidate their family assets and improve their food security directly – by means of increased production for self-consumption – and not through market mediation.

3. **Brazilian family farmers and public policies**

Large export-oriented farms have always been prominent actors in Brazilian agriculture, being the target of most Federal agricultural policies until the 1990s (Deininger and Byerlee, 2012; Rada et al., 2019). Although family farms represent 77% of the total Brazilian establishments (IBGE, 2020), they have not received enough attention in the Brazilian policymaking process. However, in the last decades, the pressure exerted by social movements, international non-governmental organisations, and the scientific community (e.g. Mattei, 2014) has prompted the Federal Government to introduce dedicated programmes for family farmers.

The establishment of a specific program for rural credit targeting family farms dates to 1995, when the Federal Government introduced the Pronaf (Ghinoi et al., 2018). Because of its longevity, this program has been re-modelled several times during its history. In the first period (1995-2002), its focus was on the improvement of credit facilities and the definition of a set of regulations aimed at facilitating the inclusion of those farmers that had historically been excluded from the benefits of the Brazilian agricultural policies (Schneider, 2006). In the second period (2003-2014), a larger amount of resources from the federal budget was allocated to this program, and access rules were modified to enlarge the number of beneficiaries and simplifying the funding system (Mattei, 2012). The third period is continuing to date: since 2014, the program has seen a reduction in the resources and number of contracts with farmers (Wesz Junior, 2020), which is related to a reformulation of the Federal policies as a consequence of the 2008 economic crisis and the changing national political landscape.

The Brazilian policies for technical assistance and rural extension (ATER) were introduced in 1975 with the creation of the Brazilian Enterprise for Technical Assistance and Rural Extension (*Empresa Brasileira de Assistência Técnica e Extensão Rural* – Embrater), a Federal agency whose aim was to facilitate the diffu-sion of new technologies in agriculture. However, Embrater was dismissed in 1992, and its activities were transferred to the State Governments (Sette and Ekboir, 2013). According to Diesel et al. (2015, p. 108), "since the beginning of the 1990s, for almost fifteen years, the Federal Government's investments in technical assistance and rural extension were not significant, until their resumption by the first Government of Luís Inácio Lula da Silva (2003-2006)"[1]. Today, both Federal and State policies are accessible by family farmers.

The programs dedicated to seeds distribution (Seeds) have always been implemented and managed by the State Governments, through the acquisition and distribution of seed stocks at a lower price compared to the market price. For example, the Paraíba Government has promoted the use of creole seeds to strengthen traditional food production for self-consumption (Petersen et al., 2013). Instead, in Rio Grande do Sul, the program has been conceived as a tool to promote farm specialization, thus fostering mechanization and the use of high-technology inputs (Silva et al., 2013); this approach has constrained farmers' strategies, reinforcing the idea of producing for the market and reducing production for self-consumption. In 2011, the Federal Government included all the State programs within the Brazil Without Misery Plan (*Plan Brasil Sem Miséria*) (Cunha, 2013), and, in 2015, it established the National Program for Seeds and Seedlings for Family Farming (*Programa Nacional de Sementes e Mudas para a Agricultura Familiar*).

4. **Data and methodology**

4.1. **The Brazilian National Household Sample Survey (PNAD)**

This paper uses data from the PNAD, which is carried out yearly by the IBGE and provides information on the socio-economic characteristics of the Brazilian population. The PNAD sample of households is generated using a three-stage probabilistic distribution and is representative of the Brazilian population at different levels of aggregation (IBGE, 2020). We use the 2014 PNAD wave (362,627 observations) because it is the only one released to date that contains data on the recipients for Pronaf, ATER, and Seeds.

The PNAD consists of eleven sections. We focus primarily on the second section, which explores the material conditions of the household unit, the fourth and fifth sections, which contain questions on the socio-demographic characteristics of the respondents, and the seventh section, which is dedicated to working activities and earnings (salaries and rents).

---

[1] Our translation from the original text (in Portuguese).

4.2. **The unit of observation: the Brazilian family farm**

According to the Brazilian Federal Law n. 11,326 of July 24th, 2006, to be recognized as a family farmer it is necessary to meet the following criteria: owning a farming area below 4 *módulos fiscais* (a unit of measurement in hectares); using mainly family labour in conducting and managing farming activities; and having a minimum income from these activities. In operational terms, the Brazilian policies use the definition of the National Register for Family Farming (*Cadastro Nacional da Agricultura Familiar*), which integrate the above requirements by specifying that at least 50% of the workforce dedicated to farming activities must come from family members and at least 50% of the household income should derive from these activities. In the case of Pronaf, there is also a ceiling for the annual family gross income, i.e. R$ 360,000.

Since the PNAD dataset does not include any information on whether the households surveyed are considered family farms according to the Federal law, we identified these actors by means of filtering variables. As it is not possible to assess the engagement of family members in farming activities, we retained the households who declared to be entrepreneurs and did not hire any labour, or employed a maximum of two individuals (which was a valid criterion for accessing Pronaf until the early 2000s), or declared to be employed without any remuneration. Regarding the farm area, we retained the households owning or leasing less than a certain area of land, with the threshold set based on the State where the farms is located. The annual household gross income was used as an additional criterion to filter family farmers, using R$ 360,000 as a threshold. These three criteria had to be met simultaneously for a household to be retained in the dataset. Through this filtering step, we turned a dataset of Brazilian households into a dataset of Brazilian family farms, for a total of 17,196 observations.

4.3. **Methodology**

Our variable of interest (thereinafter "outcome variable") is whether the family farms had commercialised part of their agricultural production in the agricultural year 2013-2014. The 2014 PNAD included a question on this aspect, which has been used to generate a dummy variable[2]. The respondents who did not answer the question were excluded from the analysis, resulting in a dataset of 6,699 family farms. The PNAD does not distinguish between commercialisation through formal markets and "informal" commercialisation (e.g. though social networks), although the typology of the main purchaser is enquired in another question. We purposely include all types of commercialisation because, although the World Bank promotes family farms' integration in the markets rather than "informal" sales, the latter practice has a positive and even larger impact on the incomes and food security of the rural population, as detected in Brazil (Schneider and Niederle, 2010) and in other World regions (Varga, 2020).

Participation in the Brazilian programmes Pronaf, ATER, and Seeds are our "treatments". The 2014 PNAD included specific questions for detecting if family farms received credit from Pronaf, technical assistance support, and/or accessed seeds distribution. For ATER and Seeds, family farms had the possibility to access different Federal and State programmes providing these services. However, we considered ATER and Seeds as two unique macro-policies because the aims of the Federal Government and State Governments did not differ.

As illustrated by Blundell and Costa Dias (2000; 2009), evaluating a policy intervention in a non-experimental context requires that the policy effect for each actor (the "treatment effect") must be independent from the effects on other actors. This condition is fundamental to assess average effects on treated populations (Rubin, 1980). Furthermore, information is available for treated and untreated actors, i.e. those who benefited from a policy intervention and those who did not. However, the impact of a policy is measured by the difference in the outcome variable for the treated with and without the treatment (i.e. the counterfactual), which it is technically impossible to estimate because an actor cannot present two states at the same time. Another issue specifically related to this study is the presence of different policy interventions affecting the same family farm. The combination of these interventions, deliberate or not, may lead to synergies or con-flicts that influence the desired outcome (Howlett and Rayner, 2013; Cunningham et al., 2016). Hence, we need to consider that family farms can have benefited from one or more policies, and we need to estimate their effects while isolating possible interactions that can lead to biased results.

---

[2] The question is: "During the period from September 28th, 2013 to September 27th, 2014, have you sold any part of the main production of this activity?". The outcome variable is equal to one if the family farmer sold part of their food production, zero otherwise. Because of the dichotomous nature of this variable, the outcome for recipient and non-recipient of Pronaf, ATER, and Seeds is a rate, i.e. the difference in the proportion of those who have sold part of the production between recipient and non-recipient (as in Guerzoni and Raiteri, 2015).

To address the above issues, we follow the approach proposed by Guerzoni and Raiteri (2015) for tackling the confounding problems arising when policy interactions are not considered. Their methodology is suitable for this study also because of the cross-sectional nature of the PNAD data.

Figure 1: Illustration of the intersections between policy programmes and policy mixes (1 = Pronaf; 2 = ATER; 3 = Seeds policy).

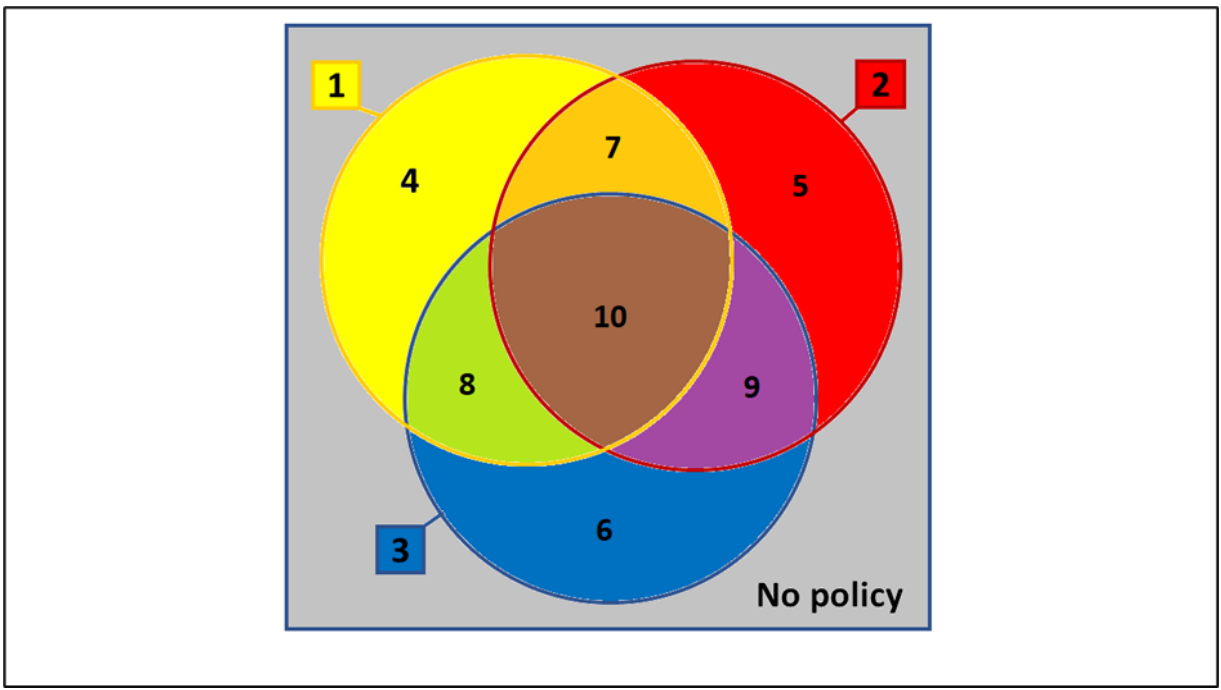

First of all, we defined the following treatments (identified by dummy variables defined according to the intersections illustrated in Figure 1):

1. Pronaf (total): 1 (dummy equal to 1) *vs* ~1 (dummy equal to 0).
2. Technical assistance (total): 2 *vs* ~2.
3. Seeds (total): 3 *vs* ~3.
4. Pronaf *only*: 4 *vs* No policy.
5. Technical assistance *only*: 5 *vs* No policy.

6. Seeds *only*: 6 *vs* No policy.
7. Pronaf & Technical assistance: 7 (1 ∩ 2) *vs* No policy.
8. Pronaf & Seeds: 8 (1 ∩ 3) *vs* No Policy.
9. Technical assistance & Seeds: 9 (2 ∩ 3) *vs* No Policy.
10. All policies: 10 (1 ∩ 2 ∩ 3) *vs* No Policy.

The first three treatments do not take account of other policy interventions simultaneously received by the family farms, and therefore are used to check for the presence of confounding effects which become visible when using treatments 4-6 (Guerzoni and Raiteri, 2015). Cases 7-10 are dedicated to the evaluation of different policy mixes.

Participation in any of the three policies is identified by means of variables inquiring whether the respondent received any type of farm credit, technical assistance, or seeds. However, we are interested in the impact of public policies, not of similar types of services obtained in the market. Considering the farm credit or technical assistance obtained from private enterprises as additional treatments would have increased hugely the number of treatments due to the potential policy mixes, thus fragmenting the sample, while ignoring them would have resulted in omitted variables and related confounding effects. For these reasons, the family farms who declared to have received agricultural credit not from Pronaf or technical assistance from private enterprises were excluded from the dataset, resulting in a sample size of 6,170 units.

Because participation in the different policy programs is not randomized, it is important to consider the presence of selection bias (Heckman et al., 1996). The self-selected group of family farms who have benefitted from one or more policies can be inherently different from those who have not received any support. Propensity Score Matching (PSM) allows to create a control group from independent untreated observations sharing similar characteristics with the treated (Rosenbaum and Rubin, 1983). For each observation, an index called Propensity Score is estimated as the probability to be treated given a set of covariates. We identified in the literature a set of variables influencing the probability of family farms to be treated, i.e. to access rural household-dedicated programs: age, gender, farm size, race, education, household size, assets, farm and off-farm incomes, and geographical location (Luan and Bauer, 2016; Fitz, 2018). The PNAD dataset includes variables for each of the above elements. For the race, we created a dummy indicating whether the respondent identified himself as "white"; for the assets, we created dummies indicating the ownership of a mobile phone, a means of transport, and access to internet; off-farm and off-farm incomes were considered separately; and for the location we included dummies for the five Brazilian macro-regions.

Farm size, usually computed as land area, is a key variable in determining the propensity of family farms to engage in commercialisation. Therefore, after implementing the analysis on the full sample, we divided it into two subsamples (smaller and larger than 2 hectares) and tested the impact of all the treatments on each group. The threshold of 2 hectares is one of those used to identify small farms at international level (European Commission, 2011, p. 5), and is also adopted by the World Bank (2003) to define smallholders. According to the 2017 Agricultural Census, in Brazil the average size of the family farms (identified based on the definition of Federal Law n. 11,326) is 20.8 hectares[3] (IBGE, 2020), and thus a family farm of more than 2 hectares is still small. Nevertheless, this threshold allows to consider the Brazilian case within the international context and to include sufficient observations in both subgroups. Our sample includes 5,512 family farms with less than 2 hectares, and 658 family farms with more than 2 hectares.

The computation of the Propensity Score is based on the following model:

$$p(X_i) = \text{Prob}(T_i=1 \mid X_i) \qquad (1)$$

where $X_i$ is the above-mentioned set of covariates characterising family farm *i*, and $T_i$ is the assignment of family farm *i* to the treatment.

PSM is used for estimating the average treatment effect on the treated (ATT), defined as follows (Caliendo and Kopeinig, 2008):

$$ATT = E(Y_{(1)} \mid T=1) - E(Y_{(0)} \mid T=1) \qquad (2)$$

where $E(Y_{(1)} \mid T=1)$ corresponds to the outcome of the family farms who accessed to the policy, and $E(Y_{(0)} \mid T=1)$ is the outcome of the same family farms if they would not have benefited from the policy. Only the first term in the right side of Equation (2) is observable ($E(Y_{(1)} \mid T=1)$), while the counterfactual term ($E(Y_{(0)} \mid$

---

[3] According to the 2017 Agricultural Census, the average farm size for all types of farms is 69.2 hectares (IBGE, 2020).

T=1)) does not exist. A solution consists in creating a counterfactual using the untreated family farms (Blundell and Costa Dias, 2000; Dehejia and Wahba, 2002). For the estimation of the ATT, two conditions need to be fulfilled: the conditional independence assumption, i.e. that family farms are assigned to the treatment independent of the observed outcome and conditionally on a set of specific covariates ($X_i$); and the common support condition, i.e. that farms with similar covariates have a positive probability of being both treated and untreated (Caliendo and Kopeinig, 2008). If both hold, it is possible to produce unbiased estimates of the ATT, and Equation (2) becomes:

$$ATT = E(Y_{(1)} | T=1, p(X)) - E(Y_{(0)} | T=1, p(X)) \quad (3)$$

where $p(X)$ is the Propensity Score given the set of covariates.

The above procedure for estimating the ATT can only be applied to cases 1-3, where the same farm can be considered treated in more than one case. For cases 4-10, which account for simultaneous participation in more than one policy, the treatments are mutually exclusive. In cases 1-3, the control group is represented by the farms which did not receive that specific treatment, while in cases 4-10 the control is always represented by those which did not benefit from any of the policies.

Following Guerzoni and Raiteri (2015), Equation (2) for treatments 4-10 becomes:

$$ATT = E(Y_{(m)} | T=m) - E(Y_{(l)} | T=m) \quad (4)$$

where *m* is the treatment of interest, and *l* is the absence of treatment. This multivariate context has been explored by Imbens (2000) and Lechner (2001), who have developed the generalized propensity score – defined as the "conditional probability of receiving a particular level of treatment given the pre-treatment characteristics" (Guerzoni and Raiteri, 2015, p. 731) – to address the identification problem arising in this situation. Hence, the propensity scores have been estimated using different discrete choice models. For cases 1-3, we estimated three logistic regressions where the dependent variable takes value T=1 in the presence of treatment and T=0 otherwise; for cases 4-10, we estimated a multinomial logistic model where the dependent variable can assume eight different values: seven corresponding to the treatments, plus the baseline outcome corresponding to the non-treated condition, i.e. T=0 for the family farms which have not benefited from any policy[4].

After estimating the propensity scores, we implemented the matching procedure using three different algorithms: Kernel, Nearest-neighbour with replacement (in line with Guerzoni and Raiteri (2015), we decided to consider the three nearest neighbours), and Radius (with a radius of 0.10). This choice enabled us to compare different results as a robustness check (Ogutu et al., 2014).

To increase robustness further, we computed the standard errors of the ATT as bootstrapped standard errors using 1,000 bootstrap replicates for each treatment and for small and large family farms separately.

5. **Results and discussion**

Our final dataset includes 6,170 family farms, of which 492 accessed credit through Pronaf, 497 received public technical assistance, and 351 benefited from Federal or State programmes for accessing seeds distribution. Table 1 illustrates the characteristics of different groups of family farms in terms of the variables used for the matching. It emerges that family farms differ depending on the policy received, as well as compared to the group of farms which did not benefit from any policy.

---

[4] Given the non-ordered nature of the dependent variable, the multinomial logistic model is appropriate, as it compares the probability of each treatment outcome to the baseline.

Table 1: Socioeconomic characteristics of family farms by public policy accessed.

| Variables (5,209 observations) | Total Mean | Total St. dev. | Pronaf Mean | Pronaf St. dev. | ATER Mean | ATER St. dev. | Seeds Mean | Seeds St. dev. | No policy Mean | No policy St. dev. |
|---|---|---|---|---|---|---|---|---|---|---|
| Age (years) | 48.36 | 14.05 | 49.30 | 12.30 | 49.81 | 13.04 | 48.71 | 12.49 | 48.20 | 14.30 |
| Gender (man) | 0.85 | 0.35 | 0.92 | 0.27 | 0.88 | 0.32 | 0.83 | 0.37 | 0.85 | 0.36 |
| Farm area (sq. m.) | 11,893.36 | 50,194.90 | 16,294.55 | 63,859.27 | 12,458.92 | 11,013.24 | 16,738.67 | 91,738.69 | 11,432.11 | 49,333.99 |
| Race (white) | 0.36 | 0.48 | 0.56 | 0.50 | 0.55 | 0.50 | 0.39 | 0.49 | 0.34 | 0.47 |
| Education (years) | 5.44 | 3.92 | 6.27 | 3.75 | 6.62 | 3.87 | 5.44 | 3.97 | 5.29 | 3.90 |
| Household size | 3.52 | 1.75 | 3.29 | 1.37 | 3.48 | 1.65 | 3.94 | 1.99 | 3.52 | 1.77 |
| Mobile phone | 0.78 | 0.41 | 0.90 | 0.31 | 0.89 | 0.31 | 0.85 | 0.36 | 0.76 | 0.43 |
| Internet | 0.20 | 0.40 | 0.34 | 0.47 | 0.31 | 0.46 | 0.24 | 0.43 | 0.19 | 0.39 |
| Means of transport | 0.64 | 0.48 | 0.86 | 0.35 | 0.81 | 0.39 | 0.68 | 0.47 | 0.61 | 0.49 |
| Farm income (R$/month) | 909.38 | 1353.17 | 1540.07 | 1799.40 | 1407.70 | 1619.42 | 714.14 | 856.53 | 848.28 | 1307.81 |
| Other incomes (R$/month) | 272.46 | 659.82 | 253.83 | 435.69 | 330.84 | 781.31 | 275.71 | 696.24 | 267.70 | 652.05 |
| Macro-region North | 25.85 | - | 15.85 | - | 21.96 | - | 23.54 | - | 14.53 | - |
| - Northeast | 40.18 | - | 26.22 | - | 21.96 | - | 19.32 | - | 58.40 | - |
| - Central-West | 6.61 | - | 7.93 | - | 4.91 | - | 5.23 | - | 2.85 | - |
| - Southeast | 13.68 | - | 16.46 | - | 17.57 | - | 18.31 | - | 10.83 | - |
| - South | 13.68 | - | 33.54 | - | 33.59 | - | 33.60 | - | 13.39 | - |
| Number of obs. | 6,170 | | 492 | | 497 | | 351 | | 5,136 | |

Notes: The three policies and "no policy" sum up to more than the "total" because some farms have benefited of more than one policy, i.e. they are counted in more columns. Number of inputted observations: 14 for mobile phone, Internet, and means of transport; 501 for the main income.

While the age of the family head is similar across groups, the farms accessing Pronaf are more often managed by a white man; those receiving seeds have a larger share of female and non-white managers; and those receiving technical assistance lie in a mid-way position. White farm managers are an absolute majority among those who received support from Pronaf and ATER (60% and 51%, respectively), and a relative majority (40%) among those benefiting from Seeds. The level of education of the farm manager is lower among the farms benefiting from the Seeds policy, while the average size of the running household is larger. The ownership of a mobile phone, of a means of transport, and access to internet follow similar patterns, with the farmers receiving credit from Pronaf being better placed. The most relevant difference is observed in the average farm income, which is more than double for the family farms receiving either Pronaf of technical assistance compared to those receiving seeds, who earn on average less than those receiving no policy sup-port. Other incomes do not differ significantly across groups. Interestingly, the family farmers receiving seeds manage a larger farm – although we have no information about land productivity. Finally, the Southern macro-region is overrepresented among the farms receiving any of the policies, while almost 60% of the farms not participating in any policy come from the Northeast.

Table 2. Number and share of farms commercialising farm products (%), by policy or policy mix received.

| Policy/policy mix | All farms | | |
| --- | --- | --- | --- |
| | No. | Market (%) | Diff. [1] |
| No policies | 5,136 | 72.0% | |
| 1. Pronaf (total) | 492 | 86.6% | 13.9% *** |
| 2. Technical assistance (total) | 497 | 84.1% | 11.3% *** |
| 3. Seeds (total) | 351 | 76.6% | 3.1% |
| 4. Pronaf only | 273 | 87.2% | 15.2% *** |
| 5. Technical assistance only | 257 | 82.5% | 10.5% *** |
| 6. Seeds only | 231 | 73.2% | 1.2% |
| 7. Pronaf & Technical assistance | 153 | 87.6% | 15.6% *** |
| 8. Pronaf & Seeds | 33 | 84.8% | 12.8% |
| 9. Technical assistance & Seeds | 54 | 85.2% | 13.2% ** |
| 10. All policies | 33 | 78.8% | 6.8% |
| Total | 6,170 | 73.8% | |

Note: [1] Compared to the farms not receiving that policy for treatments 1-3, and to "No policies" for treatments 4-10. Significance level: * 10%; ** 5%; *** 1%.

Table 2 shows the distribution of the farms by policy or policy mix received, the percentage of farms that engage in commercialisation in each group, and the difference with the control group *before matching*. Overall, 74% of the family farms sell part of the production, but this figure ranges between 87.6% of those receiving both Pronaf and ATER, and 73.2% of those benefiting from Seeds and 72.0% of those receiving no policy support. The percentage of farms selling some products is significantly higher among those receiving policy support than among those receiving no support, with the only exception of those participating in the Seeds policy, exclusively or jointly with other policies (including those receiving all measures). The number of farms benefiting from each policy is quite similar, with technical assistance being slightly more prevalent than Pronaf, and Seeds around 25% less prevalent. The most common policy mix is between Pronaf and technical assistance. Results for farms below and above 2 hectares are shown in the Supplementary Material.

5.1. **Impact of single policies on family farms' commercialisation**

To assess the impact of the single policies on commercialisation, we consider each of them separately, using first regardless of whether the farms benefited simultaneously from one of the other two – treatments *Pronaf (total)* (1), *Technical assistance (total)* (2), and *Seeds (total)* (3) – and then as the only policy received – treatments *Pronaf only* (4), *Technical assistance only* (5), and *Seeds only* (6). Table 3 illustrates the results for the first group of treatments using the three different matching algorithms, Table 4 presents the results for treatments 4-6.

The percentage of farms engaging in commercialisation increases with all the three policies. However, in the case of the Seeds policy, the difference is only slightly significant when the interactions with other policies are not considered, and it becomes non-significant if this policy is analysed separately. The strongest impact on commercialisation is generated by Pronaf. The results for treatment 1 indicate a higher probability, by between 12.7% and 13.0%, that the family farms receiving credit commercialise their output; a similar, slightly larger effect is observed for treatment 4, with between 11.9% and 14.6% more farmers engaging in commercialisation compared to the baseline depending on the matching method. The impact of Pronaf is around 1.5 times stronger than ATER. The farms receiving support from ATER show a probability to commercialise which is between 8.9% and 9.7% higher than the respective control groups. Finally, the results obtained for Seeds are more heterogenous (the percentage of commercial farms grows between 3.8% and 6.0% when all recipients are considered and between 1.9% and 7.6% for the single policy), and marginal significant only for treatment 3.

Compared to the cases when the farms receiving other policies are included among the treated (Table 3), the impact on commercialisation of the policies in isolation (Table 4) is very similar when ATER is concerned – suggesting that additional policies generate limited synergies – and slightly higher for Pronaf – anticipating that farmers accessing *this policy alone* are more likely to commercialise. Benefiting from Seeds, even if jointly with other policies, does not seem to make almost any difference for commercialisation.

Table 3: Increase in the percentage of commercial family farms among policy recipients, regardless of policy interactions (treatments 1-3).

| Interaction | Treated | Controls | Matching algorithm | ATT (common support) | | | | |
|---|---|---|---|---|---|---|---|---|
| | | | | coeff. | st. err. | z | p-value | % |
| 1. Pronaf (total) | 491 | 5,678 | Kernel | 0.130 | 0.018 | 7.40 | 0.000*** | 13.0% |
| | | | Nearest neighbour (n=3) | 0.127 | 0.028 | 4.60 | 0.000*** | 12.7% |
| | | | Radius caliper (0.10) | 0.129 | 0.018 | 7.33 | 0.000*** | 12.9% |
| 2. Technical assistance (total) | 497 | 5,673 | Kernel | 0.097 | 0.018 | 5.50 | 0.000*** | 9.7% |
| | | | Nearest neighbour (n=3) | 0.095 | 0.027 | 3.48 | 0.000*** | 9.5% |
| | | | Radius caliper (0.10) | 0.096 | 0.018 | 5.45 | 0.000*** | 9.6% |
| 3. Seeds (total) | 351 | 5,819 | Kernel | 0.038 | 0.023 | 1.70 | 0.090* | 3.8% |
| | | | Nearest neighbour (n=3) | 0.060 | 0.034 | 1.77 | 0.077* | 6.0% |
| | | | Radius caliper (0.10) | 0.038 | 0.022 | 1.68 | 0.094* | 3.8% |

*Notes*: Significance level: * 10%; ** 5%; *** 1%.

Table 4: Increase in the percentage of commercial family farms among the recipients of a single policy (treatments 4-6).

| Interaction | Treated | Controls | Matching algorithm | ATT (common support) | | | | |
|---|---|---|---|---|---|---|---|---|
| | | | | coeff. | st. err. | z | p-value | % |
| 4. Pronaf only | 273 | 5,136 | Kernel | 0.146 | 0.021 | 6.86 | 0.000 *** | 14.6% |
| | | | Nearest neighbour (n=3) | 0.119 | 0.036 | 3.33 | 0.001 *** | 11.9% |
| | | | Radius caliper (0.10) | 0.144 | 0.021 | 6.77 | 0.000 *** | 14.4% |
| 5. Technical assistance only | 257 | 5,136 | Kernel | 0.093 | 0.025 | 3.74 | 0.000 *** | 9.3% |
| | | | Nearest neighbour (n=3) | 0.089 | 0.038 | 2.35 | 0.019 ** | 8.9% |
| | | | Radius caliper (0.10) | 0.092 | 0.025 | 3.72 | 0.000 *** | 9.2% |
| 6. Seeds only | 231 | 5,136 | Kernel | 0.020 | 0.031 | 0.66 | 0.508 | 2.0% |
| | | | Nearest neighbour (n=3) | 0.076 | 0.047 | 1.64 | 0.101 | 7.6% |
| | | | Radius caliper (0.10) | 0.019 | 0.031 | 0.63 | 0.528 | 1.9% |

*Notes*: Significance level: * 10%; ** 5%; *** 1%.

These results confirm that Pronaf is a market-oriented policy for supporting family farms who want to commercialise their products. As highlighted by Grisa et al. (2014), most of the Pronaf recipients are located where the most capitalized farmers operate, i.e. in the South, Southeast, and Central-West regions. Indeed, the characteristics of this policy make it appealing for family farms engaged in monoculture productions and strongly dependent on the large multinational companies operating in the Brazilian agri-food system (Ploeg, 2008; Aquino and Schneider, 2011; Niederle and Wesz Junior, 2018).

The farms benefiting from ATER are less likely to sell their products than those who accessed Pronaf. In general, ATER aims to achieve a better integration of different farming activities and does not directly affect production decisions as in the case of Pronaf. Therefore, self-consumption is seen by its recipients as an additional, feasible strategy for achieving food security. Indeed, the set of policies under this umbrella allow for several initiatives as they offers training for establishing vegetable gardens, orchards, non-commercial crops (e.g. potatoes, cassava, beans, peanuts), and small-scale breeding (such as poultry, goats, and pigs) (Grisa et al., 2010).

In the case of Seeds, the lower probability for its beneficiaries to be market-oriented is due to three main reasons: (i) its public profile as a policy mainly addressing the needs of poor farmers; (ii) the greater presence of women among its recipients, who are central for the preservation and the valorisation of production for self-consumption (Grisa et al., 2010); (iii) the characteristics of the program, which requires a pay-off in seeds instead of money.

We also assess the impact of the single policy treatments on the two sub-groups of farms identified according to the 2 hectares threshold defined in the Methodology section. Like in the whole sample, all policies have a positive impact on commercialisation. However, Pronaf has a larger impact on farms of more than 2 hectares, technical assistance on farms of less than 2 hectares. Finally, seeds distribution does not increase commercialisation for either of the groups, thus the nature of the three policies detailed above is confirmed. This is also in line with a report published by IPEA (2013), which points out that production for self-consumption is more relevant for farms with a smaller area, although larger units still maintain this practice. Detailed results are reported in the Supplementary Material.

5.2. **Impact of the policy mix on commercialisation**

As a second step, we investigate the effects of the policy mix (treatments 7-10). As shown in Table 5, 153 family farms benefited from both Pronaf and ATER, 33 from Pronaf and the Seeds policy, 54 from ATER and the Seeds policy, and 33 from the three policies together. Thus, 4.4% of all farms accessed more than one policy, while 12.3% benefited from only one of them (compared to 5.3% and 15.0% among large farms only).

Table 5: Increase in the percentage of commercial family farms among the recipients of each policy mix (treatments 7-10).

| Interaction | Treated | Controls | Matching algorithm | ATT (common support) | | | | |
|---|---|---|---|---|---|---|---|---|
| | | | | coeff. | st. err. | z | p-value | % |
| 7. Pronaf & Technical assistance | 153 | 5,136 | Kernel | 0.151 | 0.028 | 5.43 | 0.000 *** | 15.1% |
| | | | Nearest neighbour (n=3) | 0.139 | 0.046 | 3.05 | 0.002 *** | 13.9% |
| | | | Radius caliper (0.10) | 0.152 | 0.028 | 5.43 | 0.000 *** | 15.2% |
| 8. Pronaf & Seeds | 33 | 5,136 | Kernel | 0.130 | 0.062 | 2.12 | 0.034 ** | 13.0% |
| | | | Nearest neighbour (n=3) | 0.121 | 0.097 | 1.25 | 0.210 | 12.1% |
| | | | Radius caliper (0.10) | 0.130 | 0.062 | 2.11 | 0.035 ** | 13.0% |
| 9. Technical assistance & Seeds | 54 | 5,136 | Kernel | 0.132 | 0.049 | 2.68 | 0.007 *** | 13.2% |
| | | | Nearest neighbour (n=3) | 0.123 | 0.081 | 1.53 | 0.126 | 12.3% |
| | | | Radius caliper (0.10) | 0.132 | 0.049 | 2.68 | 0.007 *** | 13.2% |
| 10. All policies | 33 | 5,136 | Kernel | 0.070 | 0.070 | 0.99 | 0.322 | 7.0% |
| | | | Nearest neighbour (n=3) | 0.111 | 0.110 | 1.01 | 0.311 | 11.1% |
| | | | Radius caliper (0.10) | 0.069 | 0.070 | 0.99 | 0.324 | 6.9% |

*Notes*: Significance level: * 10%; ** 5%; *** 1%.

Table 5 shows the effect of different policy mixes on commercialisation. As well as the single policies, the effect is always positive, although that of the three-policy mix is not statistically significant. The mix between Pronaf and ATER yields a relatively larger effect, as it increases the percentage of commercial farms by 13.9% to 15.1% depending on the matching algorithm, while the mixes between Seeds and either Pronaf or ATER have an effect between 12.1% and 13.2%. These effects are slightly larger than for the policies considered separately, but smaller than their sum. Moreover, some mixes generate positive synergies (ATER slightly increases the impact of Pronaf; ATER and Seeds are much more effective when delivered jointly), while others can be counterproductive for commercialisation (Seeds in addition to Pronaf reduces the impact of the latter). This is likely due to the nature of the farms accessing each type of mix, and to the different sizes of the treated groups, despite the bias having been reduced through the matching procedure.

Finally, we assessed the impact of different policy mixes on farms above and below 2 hectares separately. Detailed results are reported in the Supplementary Material. For some treatments it was not possible to estimate the impact for farms larger than 2 hectares due to the small number of treated observations. The only positive and statistically significant impact for larger farms is observed when Pronaf and ATER are accessed jointly, although this is smaller than the impact of Pronaf alone. Focusing on farms below 2 hectares, all impacts that are significant (thus, with the only exception of the three-policy mix) are also larger than those of single policies, with the mix between Pronaf and ATER having the largest impact. Thus, appropriate policy mixes can increase the commercialisation among small farms more than the single policies, although, like in the whole sample, the effects of the mixes are smaller than the sum of the effects of the single policies.

6. **Conclusions and policy implications**

Family farming is increasingly recognized as essential for promoting rural development, reducing poverty, and achieving environmental sustainability. Its importance in emerging country like Brazil is confirmed by the large number of family farms (IBGE, 2020), which in the last decades have been targeted by Federal and State policies with the objectives to both stimulate production for commercialisation and strengthening food security. However, it is still unclear whether their impact on family farmers' decision-making has gone more in one direction or the other. By focusing on Pronaf, ATER, and the Seeds policy, we investigated their impact on family farmers' decision to allocate food production to commercialisation.

Our results show that accessing Pronaf is associated to a higher probability to engage in commercialisation. This effect is visible both when farmers participate only into this program and when they benefit from a combination of Pronaf and other policies, and regardless of farm size (though, it is stronger for farms larger than 2 hectares). Such findings are corroborated by qualitative studies highlighting the productivity-focused approach of Pronaf (Aquino and Schneider, 2011; Grisa et al., 2014). On the other hand, accessing seeds distribution alone is less likely to stimulate commercialization; however, when this is combined with either Pronaf or ATER, the probability commercialise shows a marginal increase. The effect of ATER is positive yet smaller than Pronaf; differently from rural credit, the impact of this policy on commercialization is stronger among smaller farms.

Finally, although some policies and policy mixes yield a non-significant effect, in no cases and with any degree of significance we observe a reduced propensity towards commercialisation among policy recipients. This must not lead to interpret these policies as a potential threat for food security. Indeed, we do not assess their impact on overall farm production and thus on the amount of food which is self-consumed. Rather, our results suggest that only in the case of the Seeds policy the resources received are likely to be fully internalised, while Pronaf and ATER favour commercialisation.

These findings pose some challenges for the development and implementation of public policies to support rural households. These should contribute to the socio-economic well-being of the farmers by facilitating access to output markets and therefore enabling family autonomy regarding food production choices (Schneider, 2006). However, they should also support food security in rural areas by strengthening a diversified production for self-consumption. Purely market-oriented policies are likely to increase households' food vulnerability, while strategies focused solely on supporting self-consumption result in a reproduction of socio-economic weaknesses. We have shown that different programs generated different outcomes for different typologies of family farms. In a country oriented towards international food markets like Brazil, the challenge is to ensure that increased commercial orientation does not threaten food security, and this requires a balanced policy mix with non-contradictory goals.

Our analysis has two main limitations. First, we lack information on the amount of food produced for commercialisation. The dependent variable of our model is a dummy indicating if family farmers have commercialised part of their food production or not. Thus, we could not assess the impact of the different policies in terms of amount of food allocated to the market. Another limitation is related to the assessment of food security. Even if our results indicate a higher propensity for commercialisation when benefiting from Pronaf or ATER, it does not mean that food security is neglected in this case. Further research should be dedicated to assessing food security and evaluating how it relates to the strategies for food commercialisation of Brazilian family famers. Moreover, it would be interesting to focus on other programs that can be accessed by Brazilian family farms, such as the Food Acquisition Program, the National School Feeding Program, or the Bolsa Família Program.

**References**

Abafita, J., Atkinson, J., Kim, C. S., 2016. Smallholder commercialization in Ethiopia: market orientation and participation. International Food Research Journal 23(4), 1797.

Alexander, K. S., Parry, L., Thammavong, P., Sacklokham, S., Pasouvang, S., Connell, J. G., Jovanovic, T., Moglia, M., Larson, S., Case, P., 2018. Rice farming systems in Southern Lao PDR: Interpreting farmers' agricultural production decisions using Q methodology. Agricultural systems 160, 1-10.

Aquino, J., Schneider, S., 2011. 12 anos da política de crédito do PRONAF no Brasil (1996-2008): uma reflexão crítica. Revista De Extensão e Estudos Rurais 1(2), 309-347.

Awunyo-Vitor, D., 2015. Informal financial market in Ghana: factors influencing participation by maize farmers. Savings and Development 39(1), 37-58.

Blundell, R., Costa Dias, M., 2000. Evaluation Methods for Non-Experimental Data. Fiscal Studies 21(4), 427-468.

Blundell, R., Costa Dias, M., 2009. Alternative Approaches to Evaluation in Empirical Microeconomics. Journal of Human Resources 44(3), 565-641.

Bobojonov, I., Teuber, R., Hasanov, S., Urutyan, V., Glauben, T., 2016. Farmers' export market participation decisions in transition economies: a comparative study between Armenia and Uzbekistan. Development Studies Research 3(1), 25-35.

Buchenrieder, G., 2007. Conceptual framework for analysing structural change in agriculture and rural livelihoods. Discussion Paper, No. 113, Leibniz Institute of Agricultural Development in Central and Eastern Europe (IAMO), Halle (Saale).

Buaianin, A. M., Romeiro, A. R., Guanziroli, C., 2003. Agricultura familiar e o novo mundo rural. Sociologias – Porto Alegre 5(10), 312-347.

Caliendo, M., Kopeinig, S., 2008. Some practical guidance for the implementation of propensity score matching. Journal of Economic Surveys 22(1), 31-72.

Chen, H., Wang, J., Huang, J., 2014. Policy support, social capital, and farmers' adaptation to drought in China. Global Environmental Change 24(1), 193-202.

Cunha, F. L., 2013. Sementes da Paixão e as Políticas Públicas de Distribuição de Sementes na Paraíba (PhD Tesis). UFRRJ - Instituto de Florestas, Rio de Janeiro.

Cunningham, P., Edler, J., Flanagan, K., Larédo, P., 2016. The Innovation Policy Mix., in: Edler, J., Cunningham, P., Gök, A., Shapira, P. (Eds.), Handbook of Innovation Policy Impact. Edward Elgar Publishing, Cheltenham, pp. 505-542.

Davidova, S., 2011. Semi-Subsistence Farming: An Elusive Concept Posing Thorny Policy Questions. Journal of Agricultural Economics 62(3), 503-524.

Dehejia, R., Wahba, S., 2002. Propensity score-matching methods for non-experimental causal studies. Review of Economics and Statistics 84, 151-161.

Deininger, K., Byerlee, D., 2012. The Rise of Large Farms in Land Abundant Countries: Do They Have a Future? World Development 40(4), 701-714.

Diesel, V., Mina Dias, M., Neumann, P. S., 2015. Pnater (2004-2014): da concepção à materialização, in: Grisa, C., Schneider, S. (Eds.), Políticas públicas de desenvolvimento rural no Brasil. UFRGS, Porto Alegre, pp. 107-128.

Ellis, F., 1998. Household strategies and rural livelihood diversification. Journal of Development Studies 35(1), 1-38.

European Commission, 2011. What is a small farm? EU Agricultural Economic Briefs, Brief Nº 2 - July 2011. https://ec.europa.eu/info/sites/info/files/food-farming-fisheries/farming/documents/agri-economics-brief-02_en.pdf [accessed 25 August 2020].

FAO, 2012. Sustainable agricultural productivity growth and bridging the gap for small-family farms. Inter-agency Report to the Mexican G20 Presidency. http://www.fao.org/3/a-bt681e.pdf (accessed 28 August 2020).

FAO, 2014. Agricultura Familiar en América Latina y el Caribe: Recomendaciones de Política. http://www.fao.org/3/i3788s/i3788s.pdf (accessed 28 August 2020).

FAO, 2016. Agricultura familiar y sistemas alimentarios inclusivos para el desarrollo rural sostenible. http://www.fao.org/3/a-i6403s.pdf (accessed 28 August 2020).

Fitz, D., 2018. Evaluating the impact of market-assisted land reform in Brazil. World Development 103, 255-267.

Gazolla, M., 2004. Agricultura familiar, segurança alimentar e políticas públicas: uma análise a partir da produção para autoconsumo no território do Alto Uruguai/RS (PhD Tesis). PGDR/UFRGS, Porto Alegre.

Ghinoi, S., Wesz Junior, V. J., Piras, S., 2018. Political debates and agricultural policies: Discourse coalitions behind the creation of Brazil's Pronaf. Land Use Policy 76, 68-80.

Graeub, B. E., Chappell, M. J., Wittman, H., Ledermann, S., Kerr, R. B., Gemmill-Herren, B., 2016. The State of Family Farms in the World. World Development 87, 1-15.

Grisa, C., Gazolla, M., Schneider, S., 2010. A "produção invisível" na agricultura familiar: autoconsumo, segurança alimentar e políticas públicas de desenvolvimento rural. Agroalimentaria 16(31), 65-79.

Grisa, C., Wesz Junior, V. J., Buchweitz, V. D., 2014. Revisitando o Pronaf: velhos questionamentos, novas interpretações. Revista de Economia e Sociologia Rural 52(2), 323-346.

Guerzoni, M., Raiteri, E., 2015. Demand-side vs. supply-side technology policies: Hidden treatment and new empirical evidence on the policy mix. Research Policy 44, 726-747.

Heckman, J. J., Ichimura, H., Smith, J., Todd, P., 1996. Sources of selection bias in evaluating social programs: An interpretation of conventional measures and evidence on the effectiveness of matching as a program evaluation method. Proceedings of the National Academy of Sciences of the United States of America 93(23), 13416-13420.

Hepp, C. M., Bruun, T. B., de Neergaard, A., 2019. Transitioning towards commercial upland agriculture: A comparative study in Northern Lao PDR. NJAS-Wageningen Journal of Life Sciences 88, 57-65.

HLPE, 2013. Investing in smallholder agriculture for food security. A report by the High Level Panel of Experts on Food Security and Nutrition. http://www.fao.org/3/a-i2953e.pdf (accessed 28 August 2020).

Howlett, M., Rayner, J., 2013. Patching vs Packaging in Policy Formulation: Complementary Effects, Goodness of Fit, Degrees of Freedom, and Feasibility in Policy Portfolio Design. Annual Review of Policy Design 1(1), 1-19.

Huang, J., Wang, X., Rozelle, S., 2013. The subsidization of farming households in China's agriculture. Food Policy 41, 124-132.

IBGE – Instituto Brasileiro de Geografia e Estatística, 2020. Sistema IBGE de Recuperação Automática - SIDRA. https://sidra.ibge.gov.br/ (accessed 28 August 2020).

Imbens, G. W., 2000. The role of the propensity score in estimating dose–response functions? Biometrika 87(3), 706-710.

IPEA – Instituto de Pesquisa Econômica Aplicada, 2013. A produção para autoconsumo no Brasil: uma análise a partir do Censo Agropecuário 2006. IPEA, Brasília.

Jaleta, M., Gebremedhin, B., Hoekstra, D., 2009. Smallholder commercialisation: Processes, determinants and impacts. Discussion Paper No 18. Nairobi: ILRI – International Livestock Research Institute.

Jiren, T. S., Dorresteijn, I., Hanspach, J., Schultner, J., Bergsten, A., Manlosa, A., Jager, N., Senbeta, F., Fischer, J., 2020. Alternative discourses around the governance of food security: A case study from Ethiopia. Global Food Security 24, 100338.

Kakwani, N., Neri, M. C., Son, H. H., 2010. Linkages between pro-poor growth, social programs and labor market: The recent Brazilian experience. World Development 38(6), 881-894.

Kilelu, C. W., Klerkx, L., Leeuwis, C., 2014. How dynamics of learning are linked to innovation support services: insights from a smallholder commercialization project in Kenya. Journal of Agricultural Education and Extension 20(2), 213-232.

Kostov, P., Davidova, S., Bailey, A., Gjokaj, E., Halimi, K., 2020. Can direct payments facilitate agricultural commercialisation: Evidence from a transition country. Journal of Agricultural Economics. https://doi.org/10.1111/1477-9552.12390.

Lechner, M., 2001. Identification and estimation of causal effects of multiple treatments under the conditional independence assumption. Springer, Heidelberg.

Luan, D. X., Bauer, S., 2016. Does credit access affect household income homogeneously across different groups of credit recipients? Evidence from rural Vietnam. Journal of Rural Studies 47, 186-203.

Mariyono, J., 2019. Stepping up to market participation of smallholder agriculture in rural areas of Indonesia. Agricultural Finance Review 79(2), 255-270.

Matenga, C. R., Hichaambwa, M., 2017. Impacts of land and agricultural commercialisation on local livelihoods in Zambia: Evidence from three models. Journal of Peasant Studies 44(3), 574-593.

Mattei, L., 2012. A reforma agrária brasileira: evolução do número de famílias assentadas no período pós-redemocratização do país. Estudos Sociedade e Agricultura 20(1), 301-325.

Mattei, L., 2014. O papel e a importância da agricultura familiar no desenvolvimento rural brasileiro contemporâneo. Revista Econômica do Nordeste 45(5), 83-92.

Mavrot, C., Hadorn, S., Sager, F., 2019. Mapping the mix: Linking instruments, settings and target groups in the study of policy mixes. Research policy 48(10), 103614.

Miná Dias, M., 2007. As mudanças de direcionamento da Política Nacional de Assistência Técnica e Extensão Rural (PNATER) face ao difusionismo. Revista Oikos 18(2), 11-21.

Muriithi, B. W., Matz, J. A., 2015. Welfare effects of vegetable commercialization: Evidence from smallholder producers in Kenya. Food Policy 50, 80-91.

Mwangi, J. K., Crewett, W., 2019. The impact of irrigation on small-scale African indigenous vegetable growers' market access in peri-urban Kenya. Agricultural Water Management 212, 295-305.

Niederle, P., Wesz Junior, V. J., 2018. As novas ordens alimentares. UFRGS, Porto Alegre.

Obayelu, A. E., 2016. Cross-Country Comparison of Voucher-Based Input Schemes in Sub-Sahara Africa Agricultural Transformation: Lessons Learned and Policy Implications. Agriculturae Conspectus Scientificus 81(4), 251-267.

Ogutu, S. O., Okello, J. J., Otieno, D. J., 2014. Impact of Information and Communication Technology-Based Market Information Services on Smallholder Farm Input Use and Productivity: The Case of Kenya. World Development 64, 311-321.

Ortiz, W., Vilsmaier, U., Acevedo Osorio, Á., 2018. The diffusion of sustainable family farming practices in Colombia: an emerging sociotechnical niche? Sustainability Science 13, 829-847.

Petersen, P., Silveira, L., Dias, E., Curado, F., Santos, A., 2013. Sementes ou grãos. Lutas para desconstrução de uma falsa dicotomia. Agriculturas 10(1), 36-45.

Ploeg, J. D. van der, 2008. Camponeses e Impérios Alimentares: Lutas por Autonomia e Sustentabilidade na Era da Globizacão. UFRGS, Porto Alegre.

Poole, N. D., Chitundu, M., Msoni, R., 2013. Commercialisation: a meta-approach for agricultural development among smallholder farmers in Africa? Food Policy 41, 155-165.

Rada, N., Helfand, S., Magalhães, M., 2019. Agricultural productivity growth in Brazil: Large and small farms excel. Food Policy 84, 176-185.

Rosenbaum, P. R., Rubin, D. B., 1983. The central role of the propensity score in observational studies for causal effects. Biometrika 70, 41-55.

Rubin, D. B., 1980. Randomization analysis of experimental data: The Fisher randomization test comment. Journal of the American Statistical Association 75(371), 591-593.

Santarelli, M., Marques Vieira, L., Constantine, J., 2018. Learning from Brazil's Food and Nutrition Security Policies. https://foodfoundation.org.uk/wp-content/uploads/2018/02/Learning-from-Brazilian-Food-and-Nutrition-Security-Policies_final_clean_rev_FF.pdf (accessed 28 August 2020).

Schneider, S., 2006. A diversidade da agricultura familiar. UFRGS, Porto Alegre.

Schneider, S., Niederle, P. A., 2010. Resistance strategies and diversification of rural livelihoods: the construction of autonomy among Brazilian family farmers. Journal of Peasant Studies 37(2), 379-405.

Schure, J., Levang, P., Wiersum, K. F., 2014. Producing woodfuel for urban centers in the Democratic Republic of Congo: a path out of poverty for rural households? World Development 64, S80-S90.

Sette, C., Ekboir, J. M., 2013. An overview of rural extension in Brazil: the current situation. ILAC Working Paper No. 14. ILAC, Rome.

Sibande, L., Bailey, A., Davidova, S., 2017. The impact of farm input subsidies on maize marketing in Malawi. Food Policy 69, 190-206.

Silva, P. M., Gomes, M. C., Corrêa, L. A. V., 2013. Racionalidade e inovação tecnológica: O agricultor familiar diversificado face ao processo de decisão da escolha da cultivar de milho. Revista de la Facultad de Agronomía 112(1), 35-43.

Sinyolo, S., Mudhara, M., Wale, E., 2019. The role of social grants on commercialization among smallholder farmers in South Africa: Evidence from a continuous treatment approach. Agribusiness 35(3), 457-470.

Spielman, D. J., Byerlee, D., Alemu, D., Kelemework, D., 2010. Policies to promote cereal intensification in Ethiopia: The search for appropriate public and private roles. Food Policy 35(3), 185-194.

Suess-Reyes, J., Fuetsch, E., 2016. The future of family farming: A literature review on innovative, sustainable and succession-oriented strategies. Journal of Rural Studies 47, 117-140.

Von Braun, J., 1995. Agricultural commercialization: impacts on income and nutrition and implications for policy. Food policy 20(3), 187-202.

Von Braun, J., Kennedy, E., 1994. Agricultural Commercialization, Economic Development and Nutrition. Johns Hopkins University Press, Baltimore.

Varga, M., 2019. Resistant to change? Smallholder response to World Bank-sponsored "commercialisation" in Romania and Ukraine. Canadian Journal of Development Studies 40(4), 528-545.

Varga, M., 2020. Poverty reduction through land transfers? The World Bank's titling reforms and the making of "subsistence" agriculture. World Development 135, 105058.

Vasco, C., Tamayo, G., Griess, V., 2017. The Drivers of Market Integration Among Indigenous Peoples: Evidence From the Ecuadorian Amazon. Society and Natural Resources 30(10), 1212-1228

Villarreal, F., 2018. The inclusion of family farms. Discussion of its use rural development programs in Argentina. Mundo Agrario 19(41), e091.

Wesz Junior, V. J. 2020. O Pronaf pós-2014: intensificando a sua seletividade? Revista Grifos 30(51), 89-113.

Wiggins, S., Argwings-Kodhek, G., Leavy, J., Poulton, C., 2011. Small farm commercialisation in Africa: Reviewing the issues. Future Agricultures, Research Paper No. 23. https://assets.publishing.service.gov.uk/media/57a08ad1e5274a27b20007b3/Research_Paper23.pdf (accessed 28 August 2020).

World Bank, 2003. Reaching the rural poor: A renewed strategy for rural development. Washington, DC: World Bank. http://documents1.worldbank.org/curated/en/227421468165890144/pdf/267630REACHING0THE0RURAL0POOR0.pdf (accessed 28 August 2020).

World Bank, 2007. World Development Report 2008: Agriculture for Development. World Bank, Washington, DC.

**Supplementary Material – Effect of policies and policy mixes on small and large farms separately**

Table S1 shows the characteristics of farms larger and smaller than 2 ha in terms of the matching variables. Patterns similar to the full sample can be observed when farms with less than 2 ha are considered separately; among farms with more than 2 ha, those benefiting from Pronaf and Seeds tend to sell less often than the baseline, although the difference is not significant. However, both before and after matching, the figures for farms with more than 2 ha must be considered carefully because of the small numbers, and indeed most differences are not statistically significant.

Table S1. Number of farms and share of farms commercialising some products (%), by farm size and policy or policy mix received.

| Policy/policy mix | < 2 hectares | | | > 2 hectares | | |
|---|---|---|---|---|---|---|
| | No. | Market (%) | Diff.[1] | No. | Market (%) | Diff.[1] |
| No policies | 4,612 | 71.5% | | 524 | 76.7% | |
| 1. Pronaf (total) | 422 | 85.3% | 13.2% *** | 70 | 94.3% | 16.7% *** |
| 2. Technical assistance (total) | 412 | 83.5% | 11.2% *** | 85 | 87.1% | 8.9% * |
| 3. Seeds (total) | 333 | 76.0% | 3.1% | 18 | 88.9% | 9.8% |
| 4. Pronaf only | 237 | 85.7% | 14.2% *** | 36 | 97.2% | 20.5% *** |
| 5. Technical assistance only | 203 | 82.3% | 10.8% *** | 54 | 83.3% | 6.6% |
| 6. Seeds only | 222 | 72.5% | 1.1% | 9 | 88.9% | 12.2% |
| 7. Pronaf & Technical assistance | 127 | 86.6% | 15.1% *** | 26 | 92.3% | 15.6% * |
| 8. Pronaf & Seeds | 29 | 86.2% | 14.7% * | 4 | 75.0% | -1.7% |
| 9. Technical assistance & Seeds | 53 | 84.9% | 13.4% ** | 1 | 100.0% | 23.3% |
| 10. All policies | 29 | 75.9% | 4.4% *** | 4 | 100.0% | 23.3% |
| Total | 5,512 | 73.1% | | 658 | 79.3% | |

Note: [1] Compared to the farms not receiving that policy for treatments 1-3, and to "No policies" for treatments 4-10. Significance level: * 10%; ** 5%; *** 1%.

As a first stap, we assessed the impact of the **single policy treatments** on the two sub-groups of farms identified according to the 2 ha threshold defined in the Methodology section. Table 2S illustrates the results for each of the three policies regardless of participation in the other two policies, while Table 3S shows the impact of the single, isolated policies. The positive impact on commercialisation of all policies is confirmed, although for the Seeds policy this is not statistically significant (except treatment 3 for small farms using nearest neighbour matching). Pronaf has the strongest relative impact on commercialisation, particularly for family farms of more than 2 ha, both when the effect of simultaneous policies is included (treatment 1) and when this policy is considered in isolation (treatment 4). The finding that Pronaf has a larger impact when accessed alone than jointly with other interventions is confirmed for both sub-groups of farms, but it is particularly evident for those with more than 2 ha:

when receiving credit from Pronaf, the latter are 21.2% to 23.1% more likely to commercialise, depending on the matching algorithm used (compared to between 13.2% and 13.7% for farms with less than 2 ha). On the other hand, technical assistance is found to have a stronger impact on farms with less than 2 ha, while its impact on farms with more than 2 ha is not statistically significant. The ATT of this policy for farms below 2 ha is around 10%, i.e. small farmer recipients have a probability of commercialising that is 10% higher than non-recipients. Like for Pronaf, the increase in commercialisation is larger when this policy is accessed alone (10.1-10.9%, treatment 5) than when accessed together with other programs (8.3-9.9%, treatment 2), suggesting that market-oriented farms tend to focus on a single policy. Finally, as for the Seeds policy, we observe a positive and statistically significant effect (6.5%) only for farms below 2 ha and when interaction effects are not excluded (treatment 3). These figures suggest that this policy does not push family farm recipients towards a more market-oriented production, and that commercialisation is rather increased by other policies accessed jointly with seeds distribution.

The findings for the two sub-groups of farms are in line with those for the whole sample. Moreover, we can argue that the impact of specific policies is related to the dimension of the farm: the effects of Pronaf are amplified for farms with more than 2 ha, while technical assistance has a larger impact on farms with less than 2 ha. This, together with the finding that seeds distribution does not increase commercialisation, confirms the nature of the three policies.

Table 2S: Increase in the percentage of commercial family farms among all recipients of each policy, regardless of policy interactions (treatments 1-3), for sub-groups of farms (2 ha threshold).

| Interaction | Farm dimension | Treated | Controls | Matching algorithm | ATT (common support) | | | | % |
|---|---|---|---|---|---|---|---|---|---|
| | | | | | coeff. | st. err. | z | p-value | |
| 1. Pronaf (total) | < 2 ha | 419 | 5,090 | Kernel | 0.126 | 0.020 | 6.45 | 0.000 *** | 12.6% |
| | | | | Nearest neighbour (n=3) | 0.115 | 0.031 | 3.67 | 0.000*** | 11.5% |
| | | | | Radius caliper (0.10) | 0.125 | 0.020 | 6.33 | 0.000*** | 12.5% |
| | > 2 ha | 70 | 588 | Kernel | 0.174 | 0.034 | 5.11 | 0.000*** | 17.4% |
| | | | | Nearest neighbour (n=3) | 0.219 | 0.064 | 3.42 | 0.001*** | 21.9% |
| | | | | Radius caliper (0.10) | 0.174 | 0.035 | 5.02 | 0.000*** | 17.4% |
| 2. Technical assistance (total) | < 2 ha | 411 | 5,100 | Kernel | 0.099 | 0.020 | 4.88 | 0.000*** | 9.9% |
| | | | | Nearest neighbour (n=3) | 0.083 | 0.030 | 2.76 | 0.006*** | 8.3% |
| | | | | Radius caliper (0.10) | 0.097 | 0.020 | 4.77 | 0.000*** | 9.7% |
| | > 2 ha | 83 | 573 | Kernel | 0.070 | 0.046 | 1.50 | 0.134 | 7.0% |
| | | | | Nearest neighbour (n=3) | 0.060 | 0.066 | 0.91 | 0.362 | 6.0% |
| | | | | Radius caliper (0.10) | 0.067 | 0.047 | 1.44 | 0.149 | 6.7% |
| 3. Seeds (total) | < 2 ha | 333 | 5,179 | Kernel | 0.040 | 0.025 | 1.61 | 0.107 | 4.0% |
| | | | | Nearest neighbour (n=3) | 0.065 | 0.037 | 1.74 | 0.082* | 6.5% |
| | | | | Radius caliper (0.10) | 0.039 | 0.025 | 1.58 | 0.114 | 3.9% |
| | > 2 ha [1] | 17 | 550 | Kernel | 0.078 | 0.090 | 0.86 | 0.387 | 7.8% |
| | | | | Nearest neighbour (n=3) | 0.157 | 0.135 | 1.16 | 0.244 | 15.7% |
| | | | | Radius caliper (0.10) | 0.081 | 0.091 | 0.89 | 0.375 | 8.1% |

Notes: Significance level: * 10%; ** 5%; *** 1%. [1] Based on 750 successful bootstrap replicates; convergence not achieved for nearest neighbour and radius matching.

Table 3S: Increase in the percentage of commercial family farms among the recipients of a single policy (treatments 4-6), for sub-groups of farms (2 ha threshold).

| Interaction | Farm dimension | Treated | Controls | Matching algorithm | ATT (common support) | | | | |
|---|---|---|---|---|---|---|---|---|---|
| | | | | | coeff. | st. err. | z | p-value | % |
| 4. Pronaf only | < 2 ha | 237 | 4,612 | Kernel | 0.137 | 0.024 | 5.69 | 0.000 *** | 13.7% |
| | | | | Nearest neighbour (n=3) | 0.132 | 0.038 | 3.47 | 0.001 *** | 13.2% |
| | | | | Radius caliper (0.10) | 0.135 | 0.024 | 5.56 | 0.000 *** | 13.5% |
| | > 2 ha | 36 | 524 | Kernel | 0.212 | 0.042 | 5.07 | 0.000 *** | 21.2% |
| | | | | Nearest neighbour (n=3) | 0.231 | 0.088 | 2.63 | 0.009 *** | 23.1% |
| | | | | Radius caliper (0.10) | 0.212 | 0.043 | 4.92 | 0.000 *** | 21.2% |
| 5. Technical assistance only | < 2 ha | 203 | 4,612 | Kernel | 0.102 | 0.028 | 3.69 | 0.000 *** | 10.2% |
| | | | | Nearest neighbour (n=3) | 0.109 | 0.043 | 2.53 | 0.012 ** | 10.9% |
| | | | | Radius caliper (0.10) | 0.101 | 0.028 | 3.61 | 0.000 *** | 10.1% |
| | > 2 ha | 54 | 524 | Kernel | 0.051 | 0.060 | 0.85 | 0.398 | 5.1% |
| | | | | Nearest neighbour (n=3) | 0.019 | 0.084 | 0.23 | 0.821 | 1.9% |
| | | | | Radius caliper (0.10) | 0.046 | 0.060 | 0.77 | 0.441 | 4.6% |
| 6. Seeds only | < 2 ha | 222 | 4,612 | Kernel | 0.022 | 0.032 | 0.69 | 0.492 | 2.2% |
| | | | | Nearest neighbour (n=3) | 0.060 | 0.047 | 1.29 | 0.197 | 6.0% |
| | | | | Radius caliper (0.10) | 0.020 | 0.032 | 0.63 | 0.530 | 2.0% |
| | > 2 ha [1] | 9 | 524 | Kernel | 0.106 | 0.144 | 0.74 | 0.460 | 10.6% |
| | | | | Nearest neighbour (n=3) | - | - | - | - | - |
| | | | | Radius caliper (0.10) | - | - | - | - | - |

Notes: Significance level: * 10%; ** 5%; *** 1%. [1] Based on 750 successful bootstrap replicates; convergence not achieved for nearest neighbour and radius matching.

Finally, in Table 6 we show the **impact of different policy mixes** for the two sub-groups of farms. For some treatments it was not possible to assess the impact for farms larger than 2 ha due to the small number of treated observations. Therefore, the only positive and statistically significant impact on commercialisation for this type of farms is observed when Pronaf and technical assistance are accessed jointly (14.3% to 14.4%); this is anyway smaller than the large impact of Pronaf alone (21.2% to 23.2%). The mix between Pronaf and the Seeds policy generates even a reduction in commercialisation among farms larger than 2 hectares, although this effect is not significant. The impacts on farms larger than 2 ha of mixing technical assistance with the Seeds policy, and of accessing all the policies together, either cannot be measured (ATER-Seeds), or are non-significant (three policies). Thus, when focusing on the farms above 2 ha, either the single policies are more effective than the policy mixes in promoting commercialisation, or the results are biased due the very small numbers of observations.

In turn, for farms smaller than 2 ha, all impacts that are significant (thus, with the only exception of the three-policy mix) are also larger than the impacts of the single policies. The increase in commercialisation compared to the control group ranges from 13.4% when mixing technical assistance and the Seeds policy, to 17.8% for the mix between Pronaf and ATER (if calculated with nearest neighbour matching). Limited to significant coefficients, complementing Pronaf credit with technical assistance generates a marginal benefit between 1.1% and 4.6% (or between 4.6% and 6.9% if seen the other way around, as complementing technical assistance with credit); complementing Pronaf with Seeds yields a marginal effect of 1.2% to 1.4%; and complementing technical assistance with Seeds, an effect between 2.9% and 3.3%. Thus, appropriate policy mixes can increase the prevalence of commercialisation among farms smaller than 2 ha more than the single policies although, like in the whole sample, the effects of the mixes are always smaller than the sum of the effects of the single policies.

Table 6: Increase in the percentage of commercial family farms among the recipients of each policy mix (treatments 7-10), for sub-groups of farms (2 hectares threshold).

| Interaction | Farm dimension | Treated | Controls | Matching algorithm | ATT (common support) | | | | |
|---|---|---|---|---|---|---|---|---|---|
| | | | | | coeff. | st. err. | z | p-value | % |
| 7. Pronaf & Technical assistance | < 2 ha | 127 | 4,612 | Kernel | 0.148 | 0.031 | 4.75 | 0.000 *** | 14.8% |
| | | | | Nearest neighbour (n=3) | 0.178 | 0.053 | 3.35 | 0.001 *** | 17.8% |
| | | | | Radius caliper (0.10) | 0.149 | 0.031 | 4.73 | 0.000 *** | 14.9% |
| | > 2 ha [1] | 26 | 524 | Kernel | 0.144 | 0.061 | 2.37 | 0.018 ** | 14.4% |
| | | | | Nearest neighbour (n=3) | 0.107 | 0.103 | 1.04 | 0.300 | 10.7% |
| | | | | Radius caliper (0.10) | 0.143 | 0.063 | 2.29 | 0.022 ** | 14.3% |
| 8. Pronaf & Seeds | < 2 ha | 29 | 4,612 | Kernel | 0.149 | 0.065 | 2.29 | 0.022 ** | 14.9% |
| | | | | Nearest neighbour (n=3) | 0.103 | 0.111 | 0.93 | 0.353 | 10.3% |
| | | | | Radius caliper (0.10) | 0.149 | 0.065 | 2.29 | 0.022 ** | 14.9% |
| | > 2 ha [1,2] | 4 | 524 | Kernel | -0.112 | 0.238 | -0.47 | 0.639 | -11.2% |
| | | | | Nearest neighbour (n=3) | -0.222 | 0.348 | -0.64 | 0.523 | -22.2% |
| | | | | Radius caliper (0.10) | - | - | - | - | - |
| 9. Technical Assistance & Seeds | < 2 ha | 53 | 4,612 | Kernel | 0.135 | 0.050 | 2.71 | 0.007 *** | 13.5% |
| | | | | Nearest neighbour (n=3) | 0.138 | 0.078 | 1.78 | 0.076 * | 13.8% |
| | | | | Radius caliper (0.10) | 0.134 | 0.050 | 2.70 | 0.007 *** | 13.4% |
| | > 2 ha [3] | 1 | 524 | Kernel | - | - | - | - | - |
| | | | | Nearest neighbour (n=3) | - | - | - | - | - |
| | | | | Radius caliper (0.10) | - | - | - | - | - |
| 10. All policies | < 2 ha | 29 | 4,612 | Kernel | 0.037 | 0.081 | 0.45 | 0.651 | 3.7% |
| | | | | Nearest neighbour (n=3) | 0.119 | 0.123 | 0.97 | 0.334 | 11.9% |
| | | | | Radius caliper (0.10) | 0.037 | 0.081 | 0.45 | 0.652 | 3.7% |
| | > 2 ha [2] | 4 | 524 | Kernel | 0.162 | 0.117 | 1.38 | 0.167 | 16.2% |
| | | | | Nearest neighbour (n=3) | 0.111 | 0.140 | 0.79 | 0.427 | 11.1% |
| | | | | Radius caliper (0.10) | - | - | - | - | - |

Notes: Significance level: * 10%; ** 5%; *** 1%. [1] Based on 397 successful bootstrap replicates. [2] Convergence not achieved for radius matching. [3] Non-calculable for large farms due to the small number of treated observations.